\documentclass[10pt,conference]{IEEEtran}
\usepackage{graphicx,times,amsmath,amsfonts}
\usepackage{listings}
\begin{document}    
\author{\IEEEauthorblockN{Shahriar Shahabuddin, Janne Janhunen, Muhammet Fatih Bayramoglu,\\ Markku Juntti, \IEEEauthorrefmark{1}Amanullah Ghazi, and \IEEEauthorrefmark{1}Olli Silv$\acute{\text{e}}$n}\\
\IEEEauthorblockA{Department of Communications Engineering and Centre for Wireless Communications, University of Oulu, Finland.}\\\
\IEEEauthorrefmark{1}Department of Computer Science and Engineering, University of Oulu, Finland.
}

\title{Design of a Unified Transport Triggered Processor for LDPC/Turbo Decoder}
\maketitle
\begin{abstract}
\boldmath{This paper summarizes the design of a programmable processor with transport triggered architecture (TTA) for decoding LDPC and turbo codes. The processor architecture is designed in such a manner that it can be programmed for LDPC or turbo decoding for the purpose of internetworking and roaming between different networks. The standard trellis based maximum $a$ $posteriori$ (MAP) algorithm is used for turbo decoding. Unlike most other implementations, a supercode based sum-product algorithm is used for the check node message computation for LDPC decoding. This approach ensures the highest hardware utilization of the processor architecture for the two different algorithms. Up to our knowledge, this is the first attempt to design a TTA processor for the LDPC decoder. The processor is programmed with a high level language to meet the time-to-market requirement. The optimization techniques and the usage of the function units for both algorithms are explained in detail. The processor achieves 22.64 Mbps throughput for turbo decoding with a single iteration and 10.12 Mbps throughput for LDPC decoding with five iterations for a clock frequency of 200 MHz.}
\end{abstract}

\section{Introduction}\label{1}

The forward error correction (FEC) scheme is one of the integral parts of the wireless systems. The turbo coding scheme [1] has been adopted for the air interface standard called Long Term Evolution (LTE), that has been defined by the 3rd Generation Partnership Project (3GPP) [2]. Low-density parity-check codes (LDPC) [3] are also gaining popularity as it has been chosen for IEEE 802.11n WLAN systems [4], IEEE 802.16e WiMAX systems [5] and DVB-S2 [6]. Due to their excellent performance, the turbo and LDPC codes are the primary candidates for FEC scheme for the next generation communication systems. The roaming between WLAN and LTE systems requires a multimode FEC support. Therefore, a decoder which is able to support LDPC and turbo would be beneficial. 

The hardware designs of turbo and LDPC decoders are at a matured stage due to extensive efforts of researchers. Some of the efficient hardware implementations of turbo decoder can be found in [7] and [8] and some of the efficient hardware implementations of LDPC decoders can be found in [9] and [10] etc. Sun and Cavallaro [11] even designed multimode decoders as pure hardware designs. The hardware implementations provide high throughput, but the development time is not as rapid as processor based implementations. Besides, the hardware implementations suffers from inflexibility. Therefore, the hardware implementation of a multimode decoder might not be useful for other purposes. The design presented in this paper has the potential to be used as a detector or equalizer running on factor graphs, for example. 

The software implementations provide the required flexibility to support a multimode decoder, but requires a careful design to achieve the target throughput. Programmable accelerators, which enable software-hardware co-design method might be an attractive solution to overcome these bottlenecks.

Several application-specific instruction-set processors (ASIP) for multimode decoders with high throughput have been designed in [12], [13] and [14]. However, all of the ASIPs have been programmed with low level language which does not meet the time-to-market requirement. Besides, the utilization of the same function units for both decoding algorithms has not been described explicitly. 

The design of software and hardware together to grind out the best performance and to ensure programmability is not a straightforward task. The designer needs a very efficient tool, which can be used to design the processor easily for a particular application.

In this paper, we propose a design of a processor based on the transport triggered architecture (TTA) for turbo and LDPC decoder. TTA is a very good processor template for a programmable ASIP. The TTA based codesign environment (TCE) tool enables the designer to write an application with a high level language and design the target processor in a graphical user interface at the same time [15].

Up to our knowledge, this is the first attempt to design a TTA processor for the  LDPC decoder. The turbo decoder with TTA has been designed by Salmela $et$ $al.$ [16] and Shahabuddin $et$ $al.$ [17]. As a TTA processor can be best utilized to support different algorithms, a unified processor for turbo and LDPC decoding is the natural research direction.

The max-log-MAP algorithm is used for the component decoders of turbo decoding. The parity-check matrix of size $(M,N)$ for LDPC decoding is decomposed into $M$ rows of two state trellises or supercodes. The trellis based sum-product algorithm is used on these supercodes for check node calculation. The processor achieves 22.64 Mbps throughput for turbo decoding with a single iteration and 10.12 Mbps throughput for LDPC decoding with five iterations.

The rest of the paper is organized in the following way: In Sections \ref{2} and \ref{ldpcblock}, an overview of the turbo decoding and LDPC decoding algorithms is presented. In Section \ref{unified}, the simplification techniques and the similarities between turbo and LDPC decoding algorithm are presented. The common special function unit design is presented in Section \ref{SFU}. The processor design has been presented in Section \ref{4}. In Section \ref{5}, the throughput results and comparison with other implementations are given. The conclusion is given in Section \ref{6}.

\section{Review of Turbo Decoding} \label{2}
\subsection{Turbo Decoding}
The turbo decoder consists of two soft-input soft-output (SISO) decoders, with interleavers and de-interleavers between them as shown in Fig. 1. The inputs of the turbo decoder come from the soft demodulator, which produces the log-likelihood ratios (LLR) for the systematic bits and the parity bits. The LLRs  of the systematic bit, $\it{LuI}$ and first parity bits, $\it{LcI1}$ goes to the first SISO decoder.  The SISO decoder produces soft outputs based on  these LLRs. These soft outputs are used in the second SISO decoder as the additional information. The inputs of the second SISO decoder are the LLRs coming from the systematic bits, second parity bits denoted by $\it{LcI2}$ and output of the first SISO decoder. The LLRs of the systematic bits are scrambled this time with the same interleaving pattern used at the encoder. Similarly, the soft outputs coming from the first SISO decoder are scrambled also with the same interleaving pattern, which are used as $a$ $priori$ values for the second SISO decoder. 

\begin{figure}[h]
\includegraphics[keepaspectratio,width=1\columnwidth]{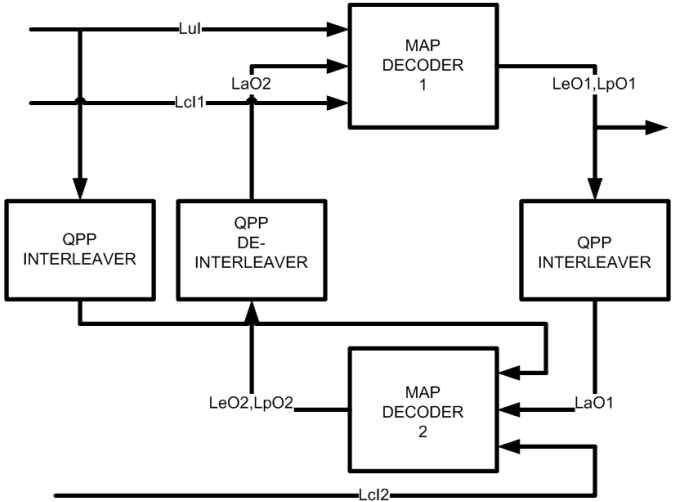}
\caption{Block diagram of the turbo decoder.}
\label{fig:}
\end{figure}
The heart of the turbo coding is the iterative decoding procedure. The output of the second SISO decoder does not produce the hard outputs immediately, but the soft output is used again in the first SISO decoder for more accurate approximation. The process continues in a similar fashion in an iterative manner.
A single iteration by both the first and the second SISO decoder is referred to as a full iteration. On the other hand, the operation performed by a single SISO decoder can be referred to as a half iteration. At the beginning of the first iteration, the $a$ $priori$ values are set at zero. Six to eight full iterations are used to achieve sufficient performance [1].

\subsection{MAP Algorithm for Component Decoder}
The MAP algorithm for the component decoder applied here has been proposed by Benedetto $\it{et}$ $\it{al}$. [18]. The algorithm can be stated like:

1.  Initialize the values of the forward state metric as $\alpha_{0}(s)=0$ if $s=S_{0}$ and $\alpha_{0}(s)=-\infty $ otherwise.

2. Calculate all the forward state metric of the same window through the forward recursion according to 
\begin{equation}
\begin{split}
\alpha_{k}(s)=\hspace{.1cm}&\max_{e}{}^{*}(\alpha_{k-1}[s^{S}(e)]+u(e)LuI[k-1]\\
&+c_1(e)LcI\it{1}[k-1]+c_2(e)LcI\it{2}[k-1]).
\end{split}
\end{equation}

3. Initialize the values of the backward state metric as $\beta_{n}(s)=0$ if $s=S_{n}$ and $\beta_{n}(s)=-\infty $ otherwise.

4. Calculate all the backward state metric of the same window through the backward recursion as
\begin{equation}
\begin{split}
\beta_{k}(s)=\hspace{.1cm}&\max_{e}{}^{*}(\beta_{k+1}[s^{E}(e)]+u(e)LuI[k+1]\\
&+c_1(e)LcI\it{1}[k+1]+c_2(e)LcI\it{2}[k+1]).
\end{split}
\end{equation}

5. The LLR values for the information and both parity bits can be calculated as following:
\begin{equation}
\begin{split}
LLR(.;O)=\hspace{.1cm}&\max_{e}{}^{*}(\alpha_{k-1}[s^{S}(e)]+c_1(e)LcI\it{1}[k-1]\\
&+c_2(e)LcI\it{2}[k-1]+\beta_{k}[s^{E}(e)]).
\end{split}
\end{equation}

For max-log-MAP algorithm, $\text{max}^*(x,y)\approx \text{max}(x,y)$ [19]. 
The decoding is done in smaller windows so that the decoding process can be done in parallel and the decoder does not have to wait for the whole block to arrive before starting the decoding process. This windowing is sometimes referred to as a sliding window method.


\section{Review of LDPC Decoding} \label{ldpcblock}
\subsection{Quasi-Cyclic LDPC Codes}
LDPC codes are linear block codes which consist of codewords 
satisfying the parity-check equation  
\begin{equation}
\mathbf{Hx}^T = \mathbf{0} \textrm{,}
\end{equation}
where $\mathbf{H}$ is the parity-check matrix and $\mathbf{x}$ is a codeword. The parity-check matrix $\mathbf{H}$ 
is `sparse' or consists of a small number of non-zero entries in case of LDPC codes.

The non-zero entries of the parity check matrix $\mathbf{H}$ are usually distributed pseudo-randomly according
to some distribution. Although this pseudo-random distribution leads to very good FER performance, it makes
the encoding and decoding of LDPC  codes difficult. Therefore, a structure is imposed on $\mathbf{H}$ to ease
the encoding and decoding by slightly sacrificing the performance. A  good trade-off between 
complexity and performance is provided by the quasi-cyclic (QC) LDPC codes [20]. 

The parity check matrix of the QC-LDPC consists of square blocks which are either all zero matrix or 
cyclic shifts of the identity matrix. This structure of the parity check matrix leads to 
efficient encoding and decoding architectures. Due to their architecture aware construction, QC-LDPC
codes have been adopted by several wireless standards such as IEEE 802.11n, IEEE 802.16e and DVB-S2.

\subsection{Decoding of LDPC Codes}

LDPC codes can be visualized by a bipartite graph consisting of check and variable nodes
which represent the rows and columns of the parity check matrix respectively. 

The decoding algorithm of LDPC codes is described as a message passing algorithm running on this graph. 
Alternative LDPC decoding algorithms differ basically in two aspects which are
message computation at the check nodes and message flow schedules. In LDPC decoding, messages
are almost always represented in LLR domain to make the decoding numerically stable. 

\begin{figure}[h]
\centering
\includegraphics[keepaspectratio,width=1\columnwidth]{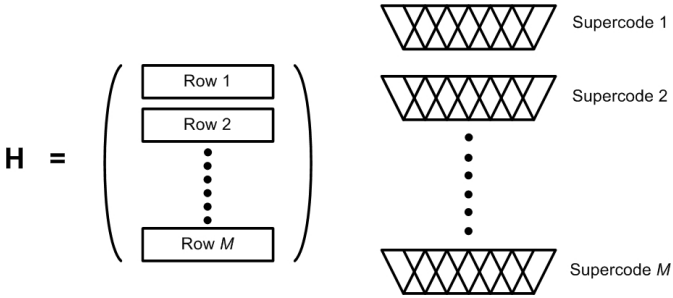}
\caption{Super codes from the parity-check matrix of LDPC codes.}
\label{fig:}
\end{figure}

The sum-product algorithm is employed at the check nodes in exact or 
approximate fashion. An approximation, which is proposed in [21],
computes outgoing messages by finding the  minimum of the absolute values of a subset of the incoming 
messages.  Although this approximation is very popular in the LDPC decoding literature, 
the hardware it imposes is not useful for turbo decoding. The hardware for computing the minimum and absolute values are redundant for turbo decoding. Therefore, using this 
approximation reduces the hardware reusability. Hence, we use the forward-backward
algorithm running on a binary trellis similar to [9] and [11]. This algorithm is derived by decomposing $(M,N)$ parity check matrix into $M$ binary or two-state trellises, which can also be called supercodes. The supercodes are shown in Fig. 2. The algorithm 
can be described for a check node of weight $Z$ as follows. 

\textbf{1.} The forward and backward recursion metrics are initialized as 
\begin{eqnarray}
  \alpha_0(0)&=&\beta_{Z}(0)=0 \nonumber \text{.} \\
  \alpha_0(1)&=&\beta_{Z}(1)=-\infty \nonumber \text{.}
\end{eqnarray}
\textbf{2.} Forward recursion metrics are computed for $k=1,2,\ldots,Z-1$ as 
\begin{eqnarray}
  \alpha_{k}(s)=\max_{b}{}^{*} \left\{  \alpha_{k-1}(s \oplus b) + (-1)^{b}Li_{k-1}  \right\} \textrm{,}
\end{eqnarray}
where $s$ and $b$ are binary variables, $\oplus$ denotes the binary addition, and $Li_k$ denotes
the incoming message from the $k^{th}$ neighbor.\\  
\textbf{3.} Backward recursion metrics are computed for \mbox{$k=Z-1,Z-2,\ldots,1$} as
\begin{eqnarray}
  \beta_{k}(s)=\max_{b}{}^{*} \left\{  \beta_{k+1}(s \oplus b) + (-1)^{b}Li_{k+1}  \right\} \textrm{.}
\end{eqnarray}
\textbf{4.} Finally the outgoing message to the $k^{th}$ neighbor is given by 
\begin{eqnarray}
  Lo_k&=&\frac{1}{2}\left(\max_{s}{}^{*} \left\{ \alpha_{k-1}(s)+\beta_{k}(s)  \right\} \right. \nonumber \\ 
  && \left.- \max_{s}{}^{*} \left\{ \alpha_{k-1}(s)+\beta_{k}(s \oplus 1)  \right\}\right)\textrm{.}
\end{eqnarray}

Notice that the steps 3 and 4 can be carried in the same run to get rid of storing $\beta_k(.)$. 

We prefer layered schedule [22] as the message flow schedule. This schedule can be formally described 
as follows. 

\textbf{1.} Initialize $A(n)$ to $\lambda(n)$ for $n=1,2,\ldots,N$ where 
$\lambda(n)$ denotes the LLR of the $n^{th}$ bit received from the channel and 
$N$ is the block length of the code. \\
\textbf{2.} For each row repeat the following. \\
\textbf{2.a} Assign $Li_k=A(\pi_j(k))-Lp_{j,k}$ where $\pi_j(k)$ denotes location 
of the $k^{th}$ $1$ on the $j^{th}$ row of $\mathbf{H}$ and $Lp_{j,k}$ is the outgoing 
message computed by $j^{th}$ row in the previous iteration for the $\pi_{j}(k)^{th}$ bit. For the first iteration 
take $Lp_{j,k}$ as $0$.  \\
\textbf{2.b.} Compute the outgoing messages according to the algorithm above for the  $j^{th}$
row.\\ 
\textbf{2.c.} Update $A(\pi_j(k))$ as $A(\pi_j(k)) \leftarrow A(\pi_j(k))+ Lo_k$ and assign $Lp_{j,k}=Lo_k$
to use in the next iteration. \\
\textbf{3.} Goto Step 2 until a certain number of iteration. \\
\textbf{4.} $A(n)$ holds the estimated LLR's  from the LDPC decoding.

\section{Shared Calculations between Turbo and LDPC Decoding} \label{unified}
There are 16 branch metric computations between two states for forward metric, backward metric and LLR calculations in the trellis diagram of an eight state convolutional code. 
\begin{figure}[h]
\centering
\includegraphics[keepaspectratio,width=.8\columnwidth]{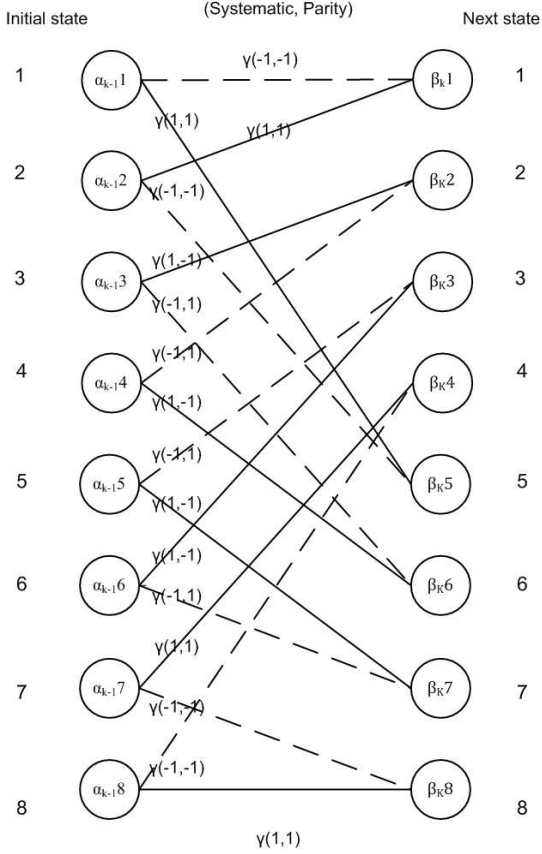}
\caption{Trellis of 3GPP turbo code.}
\label{fig:}
\end{figure}

From the trellis structure of the 3GPP turbo code in Fig. 3, it can be seen that four calculations of branch metric are being repeated to result in total sixteen calculations. The four calculations can be expressed as
\begin{equation} \small
\begin{split}
    \gamma_1 = LuI+LcI\it{1}+LcI2,\\
	 \gamma_2 = -LuI-LcI\it{1}+LcI2,\\
	 \gamma_3 = LuI+LcI\it{1}-LcI2,\\
    \gamma_4 = -LuI-LcI\it{1}-LcI2,
\end{split}
\end{equation}
where $\gamma_4$ can be represented as $-\gamma_1$ and $\gamma_3$ can be represented as $-\gamma_2$. Therefore, it is sufficient to calculate $\gamma_1$ and $\gamma_2$ only.

\subsection{Forward Metric calculation}
The trellis of the 3GPP turbo code can be divided into four butterfly pairs. Using the branch metric values given above, the forward metric calculation of a  butterfly pair can be given as
\begin{equation}
\begin{split}
\alpha_k(1) = \text{max}^*(\alpha_{k-1}(1)-\gamma_{k-1}(1),\alpha_{k-1}(2)+\gamma_{k-1}(1)) \text{.}\\
\alpha_k(5) = \text{max}^*(\alpha_{k-1}(1)+\gamma_{k-1}(1),\alpha_{k-1}(2)-\gamma_{k-1}(1)) \text{.}
\end{split}
\end{equation}

The forward metric calculation for a supercode in LDPC decoding is also similar.
\begin{equation}
\begin{split}
\alpha_k(1) = \text{max}^*(\alpha_{k-1}(0)-Li_{k-1},\alpha_{k-1}(1)+Li_{k-1}) \text{.}\\
\alpha_k(0) = \text{max}^*(\alpha_{k-1}(0)+Li_{k-1},\alpha_{k-1}(1)-Li_{k-1}) \text{.}
\end{split}
\end{equation}

The value of the branch metric in LDPC decoding equals to a single LLR value between two time instances. On the other hand, the branch metric value in turbo decoding is a combination of three LLR values.

\subsection{Backward Metric and LLR calculation}
The backward metric calculation is also similar for turbo and LDPC. For turbo decoding, the backward metric calculation of a butterfly pair can be presented as
\begin{equation}
\begin{split}
\beta_k(1) = \text{max}^*(\beta_{k+1}(1)-\gamma_{k+1}(1),\beta_{k+1}(5)+\gamma_{k+1}(1)) \text{.}\\
\beta_k(2) = \text{max}^*(\beta_{k+1}(1)+\gamma_{k+1}(1),\beta_{k+1}(5)-\gamma_{k+1}(1)).
\end{split}
\end{equation}
and for LDPC,
\begin{equation}
\begin{split}
\beta_k(1) = \text{max}^*(\beta_{k+1}(0)-Li_{k+1},\beta_{k+1}(1)+Li_{k+1}) \text{.}\\
\beta_k(0) = \text{max}^*(\beta_{k+1}(0)+Li_{k+1},\beta_{k+1}(1)-Li_{k+1})\text{.}
\end{split}
\end{equation}

The operations needed to calculate the forward and backward metric is similar. However, the output LLR computation is different for the algorithms.
The output LLR of turbo involves eight forward metric, eight backward metrics and all the branch metric in between. The calculation can be presented as

\begin{equation} \small 
\begin{split}
LLR_k=\text{max}&(\alpha_{k-1}(1)+\beta_k(1)+\gamma_1(k),\alpha_{k-1}(2)+\beta_k(5)+\gamma_1(k),\\
&\alpha_{k-1}(7)+\beta_k(8)+\gamma_1(k),\alpha_{k-1}(8)+\beta_k(4)+\gamma_1(k),\\
&\alpha_{k-1}(3)+\beta_k(6)+\gamma_2(k),\alpha_{k-1}(4)+\beta_k(2)+\gamma_2(k),\\
&\alpha_{k-1}(5)+\beta_k(3)+\gamma_2(k),\alpha_{k-1}(6)+\beta_k(7)+\gamma_2(k))\\
-\text{max}&(\alpha_{k-1}(1)+\beta_k(5)-\gamma_1(k),\alpha_{k-1}(2)+\beta_k(1)-\gamma_1(k),\\
&\alpha_{k-1}(5)+\beta_k(7)-\gamma_1(k),\alpha_{k-1}(6)+\beta_k(3)-\gamma_1(k),\\
&\alpha_{k-1}(3)+\beta_k(2)-\gamma_2(k),\alpha_{k-1}(4)+\beta_k(6)+\gamma_2(k),\\
&\alpha_{k-1}(7)+\beta_k(4)-\gamma_2(k),\alpha_{k-1}(8)+\beta_k(8)+\gamma_2(k)) \text{.}
\end{split}
\end{equation}

On the other hand, the LLR calculation of the super code in LDPC is simple.
\begin{equation}
\begin{split}
LLR_k = \frac{1}{2}(\text{max}(\alpha_{k-1}(0)+\beta_{k}(0),\alpha_{k-1}(1)+\beta_{k}(1))\\
-\text{max}(\alpha_{k-1}(0)+\beta_{k}(1),\alpha_{k-1}(1)+\beta_{k}(0))) \text{.}
\end{split}
\end{equation}

The output LLR calculation of turbo decoding needs to be divided into four parts to make the calculations similar.

\section{Special Function Unit Design} \label{SFU}

\subsection{ALPHA Special Function Unit}
A function unit can be made with three inputs and two outputs to compute the forward and backward metric. In turbo case, the unit can use $\alpha_{k-1}(1)$, $\alpha_{k-1}(2)$ and $\gamma_{k-1}(1)$ as inputs and compute the outputs of $\alpha_k(1)$ and $\alpha_k(5)$. In case of LDPC, the same unit can use $\alpha_{k-1}(0)$, $\alpha_{k-1}(1)$ and LLR as inputs and compute $\alpha_k(0)$ and $\alpha_k(1)$ as outputs. The same function unit can be used for both of the cases because the operations are same as can be seen from (9) and (10).

\begin{figure}[h]
  \centering
\includegraphics[keepaspectratio,width=.8\columnwidth]{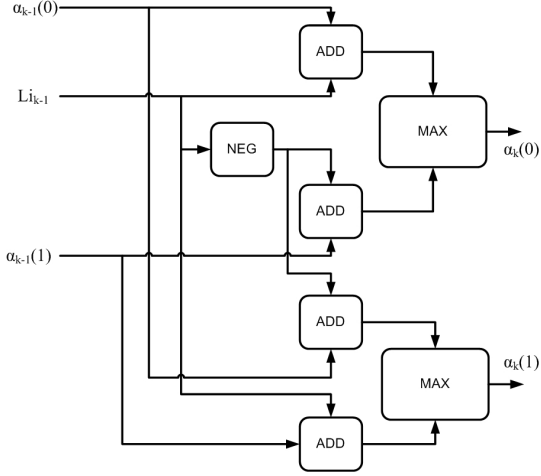}
  \caption{ALPHA unit for a single butterfly pair.}
\end{figure}

One of the ALPHA special function units calculates two forward metric values based on two earlier state forward metric in the same butterfly pair. Therefore four ALPHA unit can calculate all the necessary forward metric values for one time instant. A block diagram of the ALPHA unit used for LDPC decoding is presented in Fig. 4.

On the other hand, the LDPC can utilize these four units by processing four supercodes parallely. 

\subsection{BetaLLR Special Function Unit}
The backward metric and the LLR is computed together to reduce memory requirement. For a single algorithm it is easier to design a unit for beta separately and LLR separately. However, the LLR calcualtion of LDPC and turbo is not the same.
It can be seen from (14) that we can calculate with a special function unit the two maximization properties as
\begin{equation}
\begin{split}
output\text{\it{1}} = \text{max}(\alpha_{k-1}(0)+\beta_{k}(0),\alpha_{k-1}(1)+\beta_{k}(1)) \text{.}\\
output\text{\it{2}} = \text{max}(\alpha_{k-1}(0)+\beta_{k}(1),\alpha_{k-1}(1)+\beta_{k}(0)) \text{.}
\end{split}
\end{equation}
The unit would calculate the earlier state backward metrics $\beta_{k}(0)$ and $\beta_{k}(1)$ based on $\beta_{k+1}(0)$ and $\beta_{k+1}(1)$.  

Therefore, the BetaLLR unit takes five inputs and produce four outputs. For example, the unit takes $\beta_{k+1}(0)$, $\beta_{k+1}(1)$, $\alpha_{k-1}(0)$, $\alpha_{k-1}(1)$ and $Li_{k+1}$ as inputs and can produce the earlier state backward metrics  $\beta_{k}(0)$, $\beta_{k}(1)$, $output\text{\it{1}}$ and $output\text{\it{2}}$ of (15). 
A block diagram of the unit is given in Fig. 5.

\begin{figure}[h]
  \centering
\includegraphics[keepaspectratio,width=1\columnwidth]{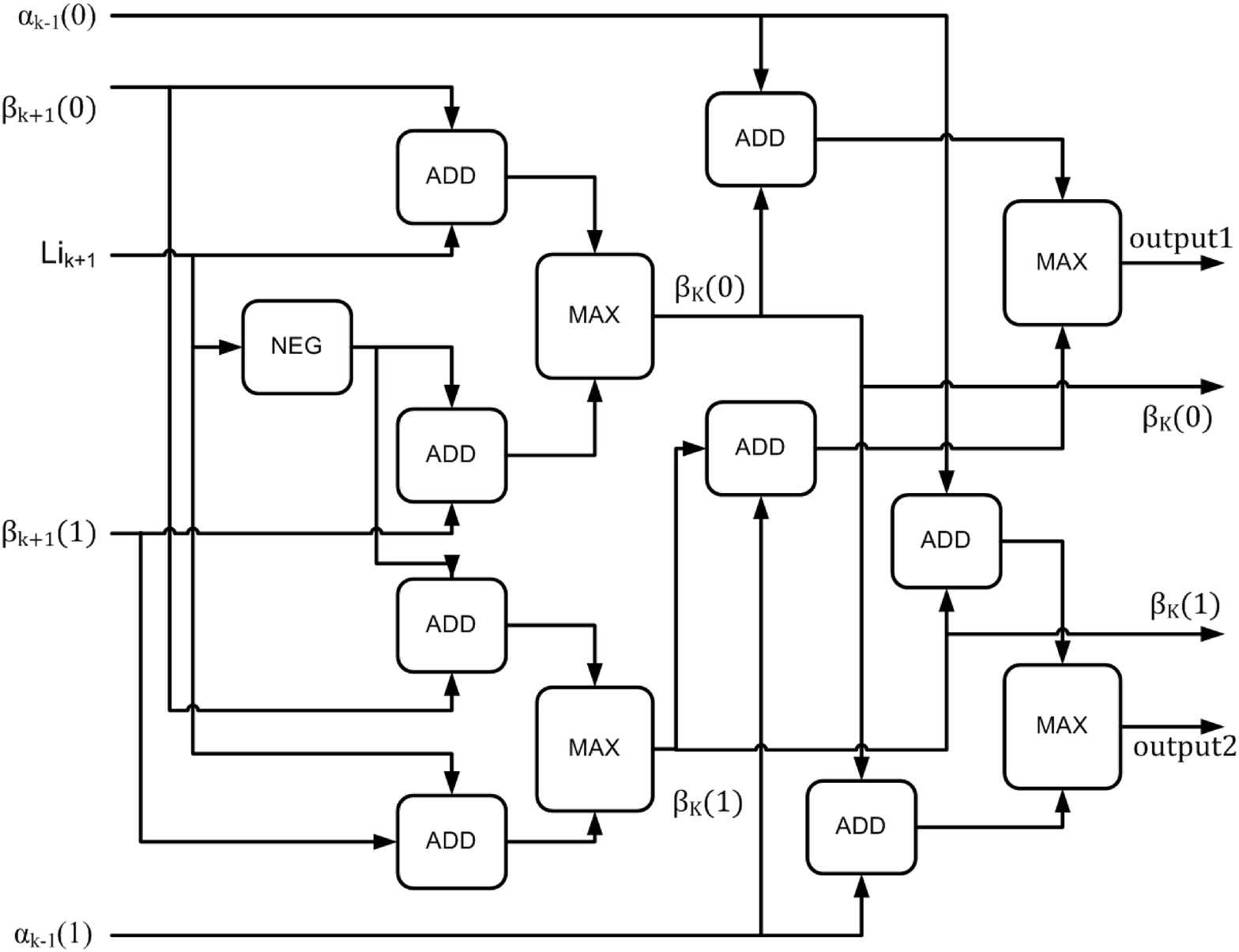}
  \caption{BetaLLR unit for a single butterfly pair.}
\end{figure}

Equation (13) has to be divided in some similar form to utilize the BetaLLR unit to compute the backward metric and output LLR for turbo decoding.

Equation (13) can be expressed as
\begin{equation} \small
\begin{split}
LLR_k=\text{max}(\text{max}(\alpha_{k-1}(1)+\beta_k(1),\alpha_{k-1}(2)+\beta_k(5))+\gamma_1(k),\\
\text{max}(\alpha_{k-1}(7)+\beta_k(8),\alpha_{k-1}(8)+\beta_k(4))+\gamma_1(k),\\
\text{max}(\alpha_{k-1}(3)+\beta_k(6),\alpha_{k-1}(4)+\beta_k(2))+\gamma_2(k),\\
\text{max}(\alpha_{k-1}(5)+\beta_k(3),\alpha_{k-1}(6)+\beta_k(7))+\gamma_2(k))\\
-\text{max}(\text{max}(\alpha_{k-1}(1)+\beta_k(5),\alpha_{k-1}(2)+\beta_k(1))-\gamma_1(k),\\
\text{max}(\alpha_{k-1}(5)+\beta_k(7),\alpha_{k-1}(6)+\beta_k(3))-\gamma_1(k),\\
\text{max}(\alpha_{k-1}(3)+\beta_k(2),\alpha_{k-1}(4)+\beta_k(6))+\gamma_2(k),\\
\text{max}(\alpha_{k-1}(7)+\beta_k(4),\alpha_{k-1}(8)+\beta_k(8)+\gamma_2(k)) \text{.}
\end{split}
\end{equation}

We can divide (16) in four parts to use the BetaLLR unit. For example, one of the parts is given here as 
\begin{equation}
\begin{split}
output\text{\it{1}} = \text{max}(\alpha_{k-1}(1)+\beta_k(1),\alpha_{k-1}(2)+\beta_k(5)) \text{.} \\
output\text{\it{2}} = \text{max}(\alpha_{k- 1}(1)+\beta_k(5),\alpha_{k-1}(2)+\beta_k(1)) \text{.}
\end{split}
\end{equation}

The branch metric $\gamma$ needs to be added or subtracted with the left side of (17) and have to use maximization unit to get the final LLR output. A block diagram is given in Fig. 6 to calculate a LLR in turbo mode.

\begin{figure}[h]
  \centering
\includegraphics[keepaspectratio,width=1\columnwidth]{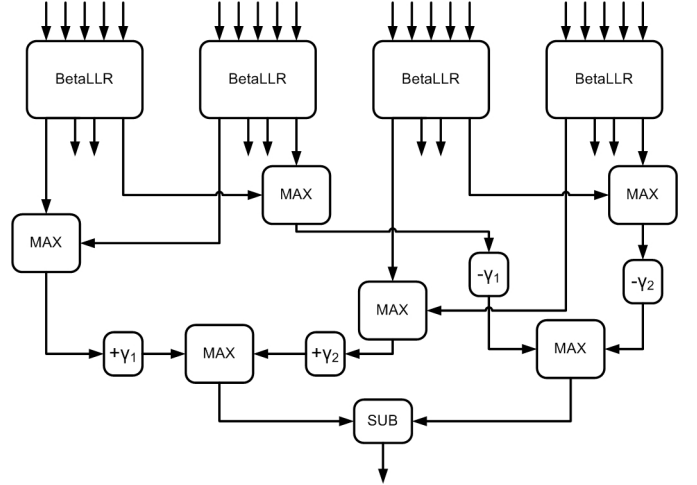}
  \caption{LLR computation with four BetaLLR units in turbo mode.}
\end{figure}

Four BetaLLR unit can be utilized in LDPC mode by processing four supercodes parallely.

\section{Transport Triggered Architecture Processor} \label{4}

\subsection{Top level architecture}
\begin{figure*}[!t]
\centering
\includegraphics[keepaspectratio,width=2\columnwidth]{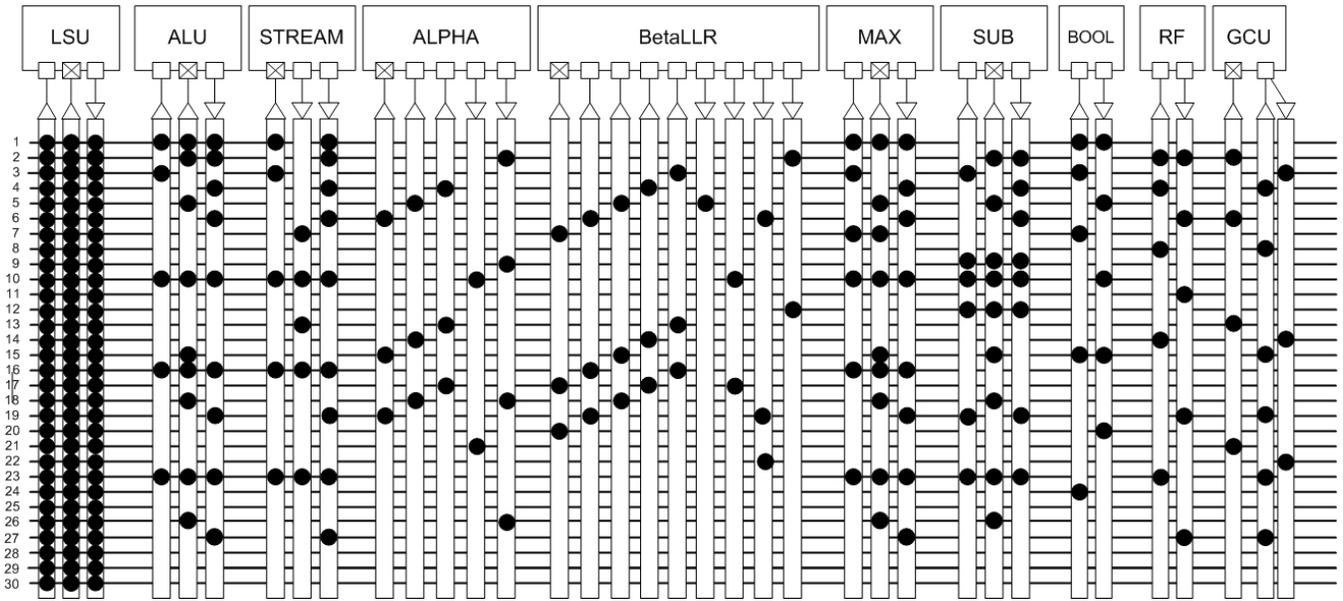}
\caption{Implemented processor with reduced number of function units.}
\label{fig:}
\vspace*{4pt}
\end{figure*}

A part of the TTA processor designed for the LDPC and turbo decoding is illustrated in Fig. 7. For readability, the whole processor figure is not given. The blocks on the upper part of the figure represent the function units and register files of the processor. The black horizontal straight lines represent the buses of the processor. The vertical rectangular blocks represent the sockets. The connection between function units and buses is illustrated by black dots in the sockets.


The fixed point processor includes load/store unit (LSU), arithmetic logic unit (ALU), global control unit (GCU) and register files. Based on the resource requirements in high level language, function units and register files are added. 

The ALU unit is used to perform the basic arithmetic operations like addition, subtraction etc. Operations like shifting right or left are also included in ALU.

For turbo decoding, the forward and backward metric computations between two states need at least four of the ALPHA and BetaLLR units. Therefore, four ALPHA and four BetaLLR units are used in the processor.   

LSU units are used to support the memory accesses. The LSU units are used to read and write memory. The memory can be read in three clock cycles and can be written in a single cycle. 

Several register files are used to save the intermediate results. In terms of the power consumption, registers can be more expensive than memory, but to meet the latency requirement register files are needed. A single Boolean register file has been included in the processor design.

Thirty buses have been used for the processor. The number of buses is crucial to ensure the parallel processing. However, the complexity also increases with the increased number of buses.

The LLR outputs have been written using a first-in-first-out (FIFO) buffer by using the function unit called STREAM. The STREAM units can write every output sample in three clock cycle.

\subsection{Programming the processor}
The processor is programmed with high level language C. Several macros have been used to call the function units and use part of that code with that specific function unit.

The turbo decoding algorithm use three blocks of 6,144 input LLRs. The blocks have been divided in smaller windows to save the memory requirements. Only the forward metrics have been saved in a window. The backward metric is calculated with the BetaLLR unit and used immediately to calculate the corresponding LLR.

The forward metric and backward metrics increase in each step and that is why the forward and backward metrics is normalized to avoid memory overflow. 

Before processing the ALPHA and BetaLLR values, the $\gamma$ values need to be calculated. As shown in figure 7, The output of the BetaLLR also needs to maximize and added or subtracted with $\gamma$ values to find the output LLR.

In the LDPC mode, the processor is programmed for the LDPC code of IEEE WLAN 802.11n of code rate 1/2 and output block size of 648. Due to the data dependencies, a single special function unit is used several times to calculate the required forward and backward metric values of a supercode in serial fashion. For example, the first row of the $\mathbf{H}$ matrix of this particular code configuration has seven nonzero elements. The two-state trellis should be a matrix of $8 \times 2$ to calculate the forward and backward metrices for this row. The initilization values of the metrices are zero. Therefore, only one ALPHA and one BetaLLR units are used seven times each to get all the necessary output LLR from this row. Four of this rows can be processed in parallel as there are four ALPHA and BetaLLR units available.

The variable node update is done by simply adding the LLR outputs of the super codes with the corresponding original LLRs of the same position. Shifting operations are required to

\section{Results and Discussion} \label{5}
The designed processor takes 166,224 clock cycles to process three blocks of 6,144 samples for a full iteration for the turbo decoding. The processor takes 10,368 clock cycles to decode a LDPC code for IEEE WLAN 802.11n of block size 648 and code rate 1/2 after five iterations.

The throughput can be calculated using the following equation as
\begin{equation}
\text{Throughput} = \frac{\text{Size of the code block} \times \text{device clock frequency}}{\text{latency} \times \text{number of iterations}}\text{.}
\end{equation}

The throughput achieved for the turbo mode is 22.64 Mbps for a single iteration and for LDPC mode 10.12 Mbps for five iterations for a clock frequency of 200 MHz. 

The buses of the processor are perfectly utilized to achieve the best possible result due to the perfect scheduling. The number of some of the operations during the algorithm execution has been summarized in Table \ref{tab1}.

\begin{table}[h]
\centering
\caption{Number of operations}
\label{tab1}
\begin{tabular}{|c|c|c|}
\hline

Operation & \# of OPS in turbo & \# of OPS in LDPC\\
\hline
ADD & 431,009 & 87,134\\
\hline
SUB & 96,354 & 14,231\\
\hline
MAX & 43,008 & 0\\
\hline
ALPHA & 24,576 & 2,376\\
\hline
BetaLLR & 24,576 & 2,376\\
\hline
STREAM & 6,144 & 648\\
\hline
\end{tabular}
\end{table} 

The number of addition operations does not only represent the addition for the algorithm, but for several other purposes like loop indexing for the code. The maximization units are not used in case of LDPC decoding because the maximization operations are done inside the ALPHA and BetaLLR units.

A comparison with different other programmable implementations of turbo decoder has been presented in Table \ref{tab2}. The throughput results are normalized for a clock frequency of 200 MHz. Our proposed processor with turbo mode provides very good throughput compared to other programmable implementations. The TTA processors of [16] and [17] provide higher throughputs but the designs were dedicated for only turbo decoding.
\begin{table}[h]
\centering
\caption{Programmable Turbo Processors}
\label{tab2}
\begin{tabular}{|c|c|c|c|}
\hline

Reference & Architecture & Algorithm & Throughput\\
\hline
[23] & TMS320C6201 DSP & max-log-MAP & 2 Mbps\\
\hline
[24] & VLIW ASIP & max-log-MAP & 5 Mbps\\
\hline
proposed & TTA proc. in turbo mode & max-log-MAP & 22.64 Mbps\\
\hline
[17] & TTA proc. for LTE & max-log-MAP & 31.21 Mbps\\
\hline
[25] & Nvidia C1060 & max-log-MAP & 33.85 Mbps\\
\hline
[16] & TTA proc. & max-log-MAP & 98 Mbps\\
\hline
\end{tabular}
\end{table} 

A comparison with different other programmable implementations of LDPC decoder has been presented in Table \ref{tab3}. The throughput results are normalized for a clock frequency of 200 MHz. Our proposed processor with LDPC mode provides moderate throughput compared to most of the programmable implementations.

\begin{table}[h]
\centering
\caption{Programmable LDPC processors}
\label{tab3}
\begin{tabular}{|p{1cm}|p{2cm}|p{2cm}|p{2cm}|}
\hline

Reference & Architecture & Algorithm & Throughput\\
\hline
[26] & TMS320C64xx DSP & min-sum & 1.8 Mbps @ 10 it.\\
\hline
proposed & TTA proc. for LDPC & supercode based sum-product & 10.12 Mbps @ 5 it.\\
\hline
[27] & SDR SODA & min-sum & 15.2 Mbps @ 10 it.\\
\hline
[14] & VLIW ASIP & offset min-sum & 53 Mbps @ 10 it.\\
\hline
[12] & VLIW ASIP & offset min-sum & 16.32 - 128.5 Mbps @ 10-20 it.\\

\hline
\end{tabular}
\end{table}

Alles $\it{et}$ $\it{al}$. presented an efficient implementation of multimode decoder in [12]. The ASIP achieved 34.5 Mbps to 257 Mbps for LDPC codes of different code configurations and block size of IEEE 802.11n when the clock frequency is 400 MHz. The lowest throughput of 34.5 Mbps at 400 MHz clock frequency after 20 iterations was achieved when the code rate and block size is the same as presented in this paper. The design of [14] achieved high throughput with a different code configuration. Besides, all the implementations presented here used the assembly language.

\section{Conclusion} \label{6}
The paper discussed the design issues of a turbo and LDPC decoder on a TTA processor. The design shows the promise of the possibility of designing several decoding techniques on a single TTA processor. The processor designed in this paper can be used for tasks beyond decoding, for instance, it can be programmed for detection and equalization algorithms running on factor graphs [28]. The target throughput of LTE can also be reached by multi-core TTA processor. The flexibility gained from that processor could provide very interesting results and would be a fruitful direction for future research.

\section{Acknowledgement} \label{con}
This research was supported by the Finnish Funding Agency for Technology and Innovation (Tekes), Renesas Mobile Europe, Nokia Siemens Networks, Elektrobit and Xilinx. 
Special thanks are due to Dr. Perttu Salmela from Tampere University of Technology and Dr. Frederik Naessens from IMEC for sharing their insights on programmable turbo and LDPC decoder implementations.

{}

\end{document}